\documentclass[]{article}
\usepackage{fullpage}
\usepackage{amsmath,amsfonts}
\usepackage{graphicx}
\usepackage[round]{natbib}  
\usepackage[colorlinks=true,citecolor=blue,linkcolor=black,urlcolor=blue]{hyperref}
\usepackage{multirow}
\usepackage{float,epsf,caption,subcaption}
\usepackage{longtable}
\usepackage{tabularx}
\usepackage{authblk}
\usepackage{setspace} 

\usepackage[english]{babel}
\usepackage{color,soul}
\usepackage{pdflscape} 

\title{Exploitation of material consolidation trade-offs in multi-tier complex supply networks}

\author[1,2]{Vinod Kumar Chauhan\footnote{Corresponding author (This work was done at University of Cambridge UK.)}}
\author[3]{Muhannad Alomari}
\author[3]{James Arney}
\author[1]{Ajith Kumar Parlikad}
\author[1]{Alexandra Brintrup}
\affil[1]{Institute for Manufacturing, University of Cambridge, UK}
\affil[2]{Department of Engineering Science, University of Oxford, UK}
\affil[3]{Data Labs, Rolls-Royce, UK}

\begin{document}
	
\maketitle
            
\begin{abstract}
While consolidation strategies form the backbone of many supply chain optimisation problems, exploitation of multi-tier material relationships through consolidation remains an understudied area, despite being a prominent feature of industries that produce complex made-to-order products. In this paper, we propose an optimisation framework for exploiting multi-to-multi relationship between tiers of a supply chain. The resulting formulation is flexible such that quantity discounts, inventory holding, and transport costs can be included. The framework introduces a new trade-off between tiers, leading to cost reductions in one tier but increased costs in the other, which helps to reduce the overall procurement cost in the supply chain. A mixed integer linear programming model is developed and tested with a range of small to large-scale test problems from aerospace manufacturing. Our comparison to benchmark results shows that there is indeed a cost trade-off between two tiers, and that its reduction can be achieved using a holistic approach to reconfiguration. Costs are decreased when second tier fixed ordering costs and the number of machining options increase. Consolidation results in reduced inventory holding costs in all scenarios. Several secondary effects such as simplified supplier selection may also be observed. 

\end{abstract}



\section{Introduction}
\label{sec_intro}
Procurement of parts from suppliers is a key task in the supply chain management, greatly impacting its competitiveness and performance (\cite{Amid2006}). In many industries, procurement cost often forms the highest proportion of total cost of a product (\cite{Willard2012}).
	
\textit{Consolidation} has been a key underlying strategy in the context of procurement decisions. Consolidation, as the name suggests, is a process of combining related activities or materials to improve performance of a supply chain resulting from cooperation and coordination (\cite{schulz2010horizontal,chadha2022freight}) and can help in reduction of costs, increase efficiency and improve performance (\cite{vaillancourt2016theoretical,Giampoldaki2023}). Material consolidation consists of purchasing, transportation and inventory activities (\cite{brauner1993consolidation}), where a buyer or a set of buyers may choose to group items or orders to obtain quantity discounts (\cite{monczka1993supply,hagberg2022consolidation}). While this helps in increased efficiency and reduction of costs it can reduce flexibility of sourcing options, thus reducing supply chain resilience. Inventory consolidation considers relocation of warehouses to increase inventories in order to make use of reduced operational costs but may introduce increased transport cost (\cite{wanke2009consolidation,ralfs2022inventory}). Transportation consolidation merges small deliveries into single dispatch of economical load but increase uncertainty in delivery times (\cite{trent1998purchasing,ccetinkaya2005coordination,Torbali2023}).
	
Consolidation activities to date have been overwhelmingly studied within the span of single supply echelons (\cite{stenius2018sustainable}). This is not surprising, as buyers often have control of their dyadic connections, gradually losing both visibility and influence beyond their immediate connections, making consolidation decisions not applicable beyond their immediate connections. There are, however, an increasing number of industrial contexts where a buyer may influence its wider supply chain, and there is willingness for cooperative decision making for collective performance (\cite{chauhan2023real}). Examples include production of complex, made-to-order products, such as heavy machinery, turbines, aerospace products, and medical devices. Due to long-term supply relations involved in these sectors, a manufacturer may be involved in configuration of whole supply chain. In addition to whole supply chain configurability, longevity of relationships necessitates de-risking through multi-sourcing activities. This increased span of control, coupled with multi-sourcing offers a unique multi-to-multi relationship structure whereby products may be consolidated further upstream, affecting cost structures at different tiers.
	
In this paper, we highlight this understudied multi-tier consolidation problem presented by the above context and formulate it through a case study. We term this new consolidation opportunity as \textit{``multi-tier material consolidation problem"}.

\begin{figure}[H]
    \centering
    \includegraphics[width=70mm]{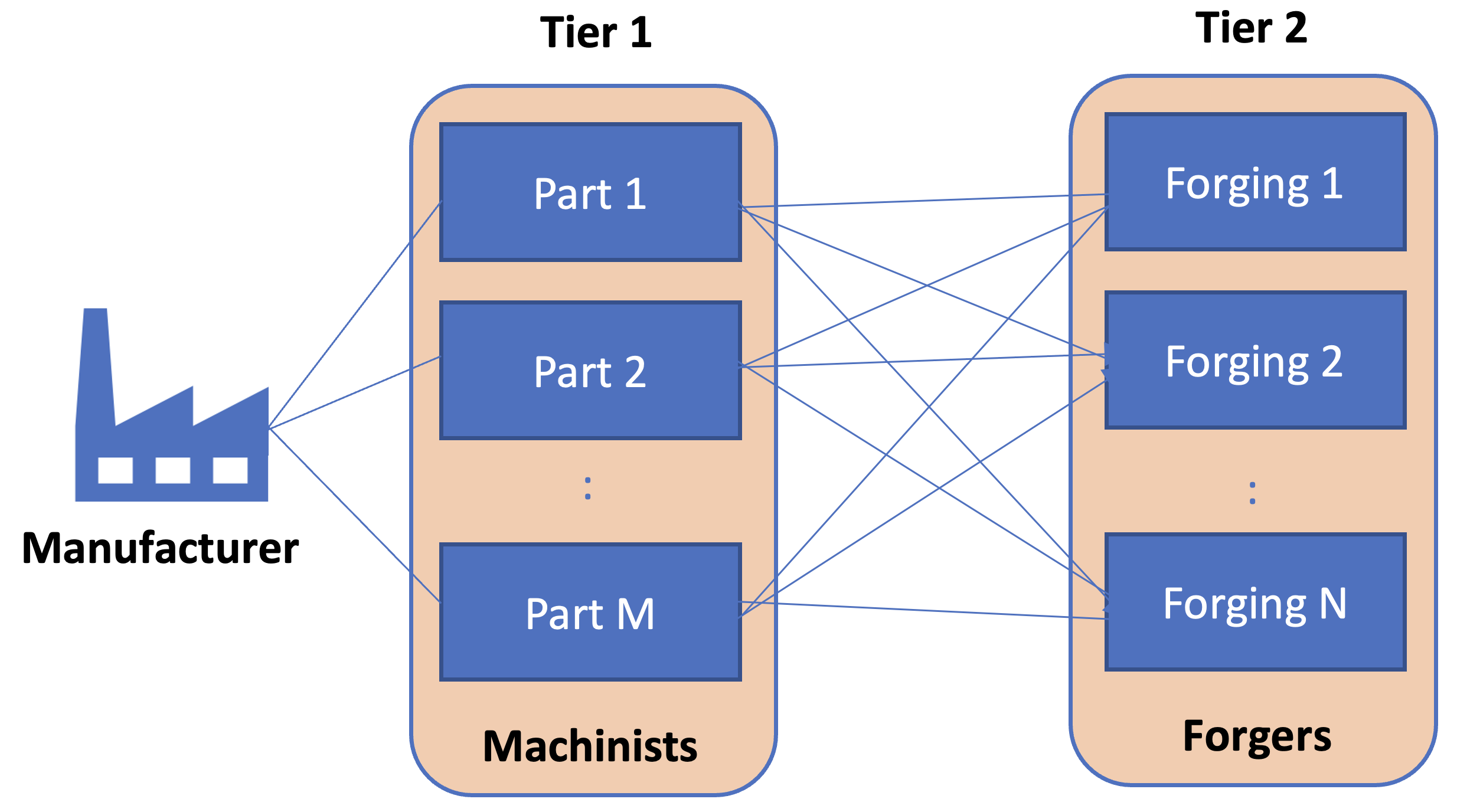}
    \caption{Two-tier supply chain of a manufacturing company}
    \label{subfig_SC}
\end{figure}

To contextualise the multi-tier consolidation problem, we consider following example from an aerospace industry (Fig.~\ref{subfig_SC}). Here, aircraft engines are produced, requiring different types of parts, which manufacturer outsources from a set of certified machining suppliers. These Tier 1 suppliers need different types of forged metal to manufacture final finished parts, which are themselves outsourced to Tier 2 forging suppliers. The forgings that could be used for manufacturing different parts is predetermined by the company.

The forging process involves manufacturing roughly shaped parts from melted alloys and machining refines those into final parts. The supply chain has N different forgings to manufacture M different parts, and creates a multi-to-multi relationship between forgings and parts. That is, one forging can be used to manufacture many parts, and similarly, one part can be manufactured in multiple ways from different forgings. The total procurement cost of parts from Tier 1 and forgings from Tier 2 depend on ordering cost, unit cost and consequent transportation costs. Since parts can be manufactured in multiple ways from different forgings, requiring different machining costs, forgings can be consolidated into a smaller set, thereby, reducing the cost of forgings at the expense of increased machining time to manufacture parts from a limited set of forgings, and hence increased machining cost. Thus, there is a trade-off between the reduced cost of forgings at Tier 2 and increased machining cost at Tier 1. Additionally, since forging process takes longer compared to machining, the company also maintains a specific inventory of forgings. This leads the company to order additional forgings resulting in extra purchasing cost and cost of holding inventories. 

\begin{figure}[htb!]
    \centering
    \includegraphics[width=70mm]{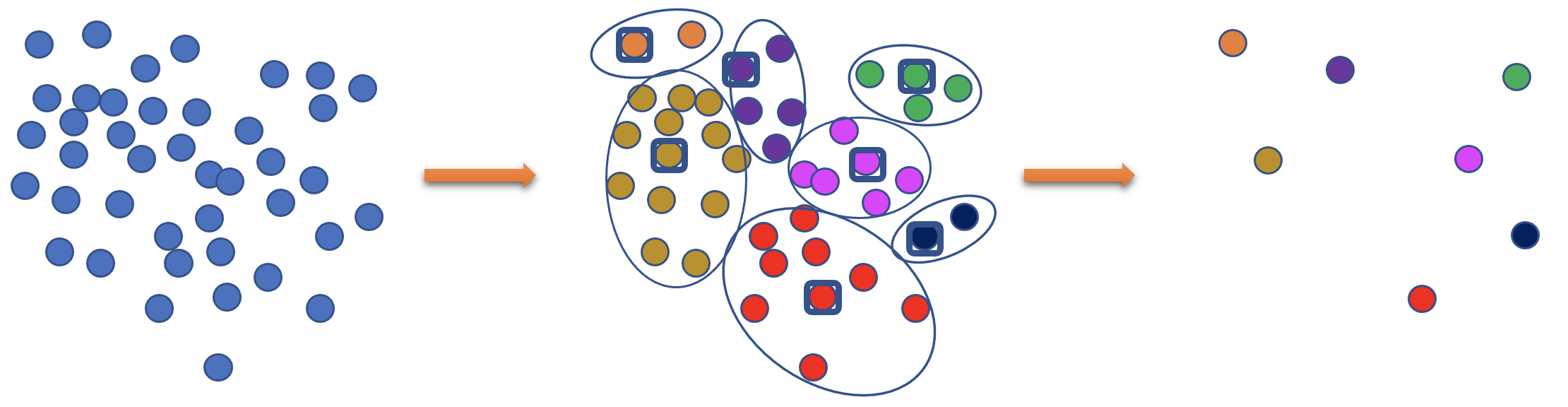}
    \caption{Illustration of the multi-tier consolidation problem: Different colours represent clusters where a square-enclosed forging is used to replace all other forgings in its cluster.}
    \label{subfig_clustering}
\end{figure}

Hence, objective of the multi-tier material consolidation problem in this example is to minimise overall procurement cost across the supply chain by consolidating forgings and striking an optimal balance between cost of forgings at second tier and machining cost at first tier. This problem can be visualised as a clustering problem, as presented in Fig.~\ref{subfig_clustering}. Here, each forging is represented as a point in some space depicted as a circle (as shown in left panel). The problem is to find clusters whereby all forging in a cluster can be replaced with a single forging from the group. That is, the selected forging is used to manufacture all machined parts that were manufactured using different forgings in the cluster (as shown in middle-panel). Thus, we need to find minimum number of clusters, and hence minimum number of forgings in the consolidated set (as shown in right-panel), which balances the trade-off between cost of forgings and machining cost. Since ordering items in different quantities affects the suppliers so quantity discounts also need to be considered.

The rest of this paper is structured as follows. In Section~\ref{sec_literature}, we present related work on supply chain consolidation to situate context of our contribution. In Section~\ref{sec_case_study}, we characterise the multi-tier material consolidation problem, formulate it using an aerospace supply chain as a case study, discuss our solution approach and choices of modelling languages as well as solver libraries, and present our analysis. We then present concluding remarks, limitations and future scope of the study in Section~\ref{sec_conclusion}.

\section{Related work}
\label{sec_literature}
Scholars in supply chain management have proposed several consolidation strategies to improve cost against operational decision criteria. These can be broadly categorised as purchasing, shipment, inventory and part consolidation (\cite{brauner1993consolidation}), as discussed below.

\subsection{Purchasing consolidation}
\label{subsec_Pconsolidation}
Purchasing consolidation considers regrouping of items for purchase, which may involve grouping of multiple, related types of products to be purchased from same supplier in order to obtain contractual and logistics discounts (\cite{monczka1993supply,chauhan2023real}), or pooling items to be purchased with other buyers (group buying) to increase economies-of-scale and obtain a reduction on unit cost of production and delivery (\cite{vaillancourt2017procurement,hu2022demand}). Both of these strategies may result in a loss of flexibility, due to the need to align production and deliveries with other product lines (in case of product grouping), or other buyers (in case of group buying) (\cite{vaillancourt2016theoretical}). Early deliveries may result in increased inventory costs (\cite{guiffrida2006cost}), and over-reliance on a single supplier may increase risk and opportunism (\cite{chopra2014reducing}). 
	
A related but separate strand of consolidation literature considers multi-sourcing decisions determining number of suppliers supplying an item. It is generally presumed that single-sourcing, i.e., procurement from a single supplier, results in the cheapest unit cost (\cite{Silbermayr2016}) while dual-sourcing and multi-sourcing avoid supplier monopoly of items and help reduce the risk of disruptions in a supply chain (\cite{Tomlin2006}).

Researchers have proposed a number of analytical models to characterise these trade-offs. For example, \cite{gaur2020impact} considered a real-world case study of an automotive-parts manufacturer to study impact of disruption on a closed-loop supply chain using sourcing policies. They developed a mixed-integer non-linear programming model for the problem and found that, under the risk of supply chain disruption, multi-sourcing generates more profit as compared to single sourcing.

\subsection{Shipment, volume and order consolidation}
\label{subsec_Sconsolidation}
Shipment consolidation, also known as freight consolidation, transportation consolidation and terminal consolidation, is a logistics strategy which refers to merging of small deliveries into a single dispatch of economical load (\cite{ulku2012dare,wagner2023better}). This helps in increased efficiency and reduction of CO$_2$ emissions, and delivery costs, e.g., \cite{munoz2021impact} investigated shipment consolidation by pooling and developed a mixed integer linear programming (MILP) model to study its effect on CO$_2$ emission, distance and transport costs. However, the practice can cause uncertainty in delivery times leading to poor service for customers (\cite{masters1980effects}).
	
Volume consolidation, i.e., a strategy where a buyer purchases most of its supply from one supplier, results in shipment consolidation and helps to reduce shipping costs. For example, \cite{Cai2010} studied volume consolidation and its effect on supply chain outcomes. Through an empirical study, they found that volume consolidation enhances buyer's ability to learn from the supplier, and supplier performance. However, coordination costs negatively affect buyer satisfaction and supplier performance.
	
Order consolidation refers to consolidation of a customer's orders at a delivery station so as to organise delivery in fewer trips. For example,
\cite{zhang2019order} studied order consolidation for last-mile split-delivery in online retailing and developed an integer programming model to study the trade-off between splitting orders and consolidating shipments.

\subsection{Inventory consolidation}
\label{subsec_Iconsolidation}
Inventory consolidation, also called facility location problem, is identification of optimal warehousing and distribution centre locations and capacities to stock up inventories with aim of meeting customer demand (\cite{wanke2009consolidation,SEYEDAN2023100024}). Minimisation of inventory holding locations in a supply chain helps to reduce operational costs, however, leads to increased distance travelled to customers and thus increased CO$_2$ and transport costs (\cite{gabler2007supply}).
	
A classical and widely studied strategy is postponement, which refers to late differentiation of products to cater to fast changing demand (\cite{zinn2019historical}), resulting in cost savings (\cite{geetha2021effective}). For a systematic review on postponement strategy refer to \cite{ferreira2018postponement,zinn2019historical} and for a review on consolidation effect and inventory portfolio analyses refer to \cite{wanke2009consolidation_b}.

\subsection{Part consolidation through product redesign}
\label{subsec_Partconsolidation}
Part consolidation is an activity that aims to drive supply chain costs down through part redesign (\cite{nie2020optimisation,kunovjanek2022additive}). Here, the assembled unit may be redesigned to contain fewer but more complex parts leading to a trade-off between increased manufacturing cost and reduced supply chain cost (\cite{knofius2019consolidating}), whilst manufacturing lead time may depend on process technology used. For example, \cite{knofius2019consolidating} observed that part consolidation through Additive Manufacturing reduced lead times but resulted in increased total costs due to loss of flexibility.

Part consolidation is widely studied with different objectives. For example, \cite{johnson2009quantifying} applied a process-based cost model to quantify effects of parts consolidation and costs on material selection choices, and \cite{Crispo2021} studied a multi-layered topology-based optimisation approach for part consolidation in Additive Manufacturing. \cite{gan2021concurrent} explored concurrent design of product and supply chain and presented a trade-off between modularity of product and sourcing flexibility in supply chain. For a detailed review on part consolidation, refer to \cite{sigmund2013topology,liu2016guidelines,gan2016concurrent}.
	
From this brief literature review, it is clear that consolidation has been studied widely in supply chains at different levels and with different perspectives.

Our work presents a unique perspective, different from the existing research, in its focus on the trade-off between two supply tiers. We consider the problem of minimising procurement cost by consolidating material through exploitation of multi-to-multi relationships in complex made-to-order products and show that consolidation in one tier results in cost savings at that tier but increased manufacturing and inventory costs in the next downstream tier, necessitating a trade-off formulation. While our work may appear similar to recent studies on part consolidation through redesign, it's important to note that, in our case, there is no redesign activity involved. Instead, we focus on the exploitation of multiple potential relationships between materials and production across two tiers (please refer to \cite{gan2016concurrent} for a review of trade-off in concurrent design of product and supply chain). Our proposed formulation is extended to incorporate a number of other considerations including shipment and purchasing consolidation so as to explore interaction between multiple consolidation strategies. We present this conceptualisation and problem formulation next.

\section{Multi-tier material consolidation}
\label{sec_case_study}
\subsection{Problem formulation}
\label{subsec_problem}
Our context consists of a two-echelon supply chain, yielding a high-value complex assembled product with long-term supplier relationships and a large number of suppliers. Typical examples include precision engineered products such as aircraft engines, medical devices, wind turbines and heavy machinery. Here, forged metal alloys are precision machined to manufacture finished parts, which are then assembled into a final product. The manufacturer who assembles the final product has overall visibility and ability to configure the whole supply chain.

As discussed earlier (Fig.~\ref{subfig_SC}), Tier~2 involves manufacturing roughly shaped parts from melted alloys, called as forgings, and Tier~1 involves machining that refines forgings into final finished parts. A single forging can be used to manufacture multiple, different parts, and similarly, one type of part can be manufactured from a variety of different forgings. However, manufacturing of a part from different forgings results in different subsequent machining costs. A myopically ideal scenario from a machining cost reduction perspective would be to have a single forged part for single machined part, as the forging brings the part to as close a shape as possible to the machined part. However, a one-to-one relationship would increase transportation costs, and result in over-reliance on the forger. Additionally, quantities per forging type would decrease at Tier~2, preventing quantity discounts. Furthermore, as forging process takes longer compared to machining process, a certain inventory of forgings must be maintained to ensure continuity of production. This leads the company to order extra forgings resulting in extra purchasing cost and cost of holding inventories.
	
\begin{figure}[htb!]
    \centering
    \includegraphics[width=0.5\linewidth]{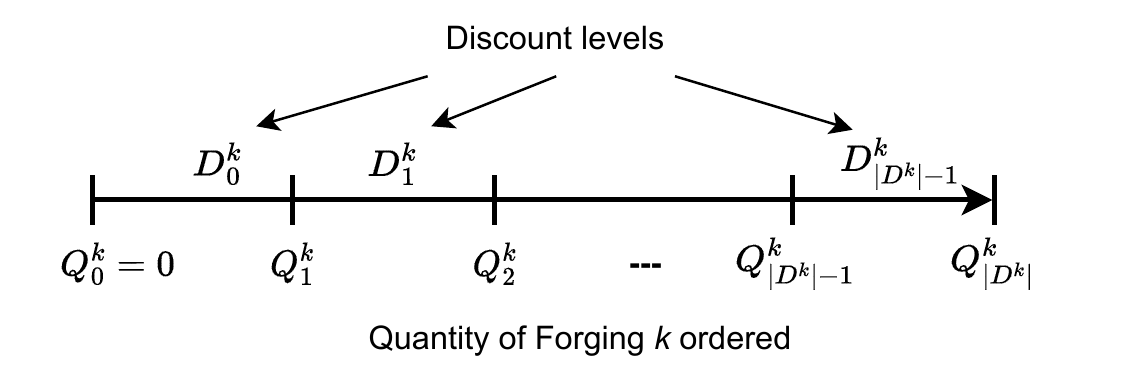}
    \caption{Quantity intervals for calculating discount levels for forging $k$. Quantity intervals and discounts for parts can be represented in a similar way.}
    \label{fig_discounts}
\end{figure}

On the other hand, consolidating forgings such that multiple machined parts can be manufactured from a single type of forging means that machining costs in Tier 1 increase, along with lead times, resulting in a trade-off between costs of the two tiers. Another myopic scenario here would include one forging creating multiple machined parts, with minimum transportation costs and maximum economies-of-scale at Tier~2, but much increased machining costs and lead time in Tier 1. To benefit from economies-of-scale, both the tiers consider discounts based on quantity of items ordered, as pictorially presented in Fig.~\ref{fig_discounts}. For a given order of parts, along with inventories, discounts on parts can be pre-computed to simplify the modelling, as information required to calculate discounts on parts is given.

Our Objective is to optimise the overall procurement cost across the supply chain for a given requirement of parts, including inventories, by consolidating Tier~2 forgings to build Tier~1 parts, which balances the trade-off between cost of the tiers; under the constraint that there should be at least one way to manufacture each part from consolidated set of forgings.

In the development of model for the problem, we assumed the following points.
\begin{itemize}
    \item[(a)] Demand is constant and a priori known for all the parts.
    \item[(b)] There is at least one way to manufacture all the parts from a given set of forgings.
    \item[(c)] There can be multiple ways to manufacture a part from different forgings but each way needs only one type of forging to manufacture the part.
    
    Given the multi-to-multi relationship between forgings and parts, and multiple manufacturing ways for a part, the forging consolidation problem can become complex due to the need to consider different forging combinations for each part. To simplify our problem, we make this assumption that can be achieved by increasing number of parts. For instance, if part P1 requires two distinct forgings, say F1 and F2, for its production, we replace P1 with two separate parts, P1A and P1B. P1A is produced using F1, and P1B is manufactured using F2. This adjustment is possible because, typically, each part relies on a single forging for its production.
    
    \item[(d)] Consolidation does not result in additional costs within the supply chain, including any costs associated with reconfiguring production for increased quantities of certain forgings. Furthermore, consolidation does not have any adverse effects on the supply chain relationships with suppliers.
        
    Consolidated forgings represent a subset of all forgings, requiring no additional machinery for production. Therefore, we anticipate minimal additional costs. Additionally, during the order assignment to suppliers stage (for details, please refer to \cite{chauhan2023real}), which follows consolidation, each supplier is assigned certain minimum orders. As a result, we expect that consolidation will not significantly impact the supply chain relationships.
   
\end{itemize}

Mathematical notations for the development of the model are defined in Table~\ref{tab_notations}.

\begin{table}[htb!]
    \centering
    \caption{Notation descriptions}
    \label{tab_notations}
    \begin{tabular}{ll}
    \hline
    \textbf{Symbols} & \textbf{Meaning} \\
    \hline
    \multicolumn{2}{l}{\textbf{Indexes:}}\\
    $i$ & index into types of parts, $i$=1,2,3,...,M \\
    $k$ & index into types of forgings, $k$=1,2,3,...,N\\
        $d$ & index into levels of discounts\\
    \hline
    \multicolumn{2}{l}{\textbf{Parameters:}}\\
    $C_F$ & total forging cost\\
    $C_M$ & total machining cost\\
    $C_I$ & total inventory cost\\
    $M_i$ & number of parts $i$ ordered\\
    $P_i$ & inventory of parts $i$ ordered\\
    $L_{ik}$ & number of forgings $k$ needed to manufacture one unit of part $i$\\
    $CMF_{i}$ & fixed ordering cost associated with part $i$\\
    $CMU_{ik}$ & per unit machining cost of part $i$ from forging $k$\\
    $CMT_{ik}$ & per unit transportation cost of part $i$ manufactured from forging $k$\\
    $CFH_k$ & per unit holding cost for forging $k$\\
    $CFF_{k}$ & fixed ordering cost associated with forging $k$\\
    $CFU_{k}$ & per unit cost associated with forging $k$\\
    $CFT_{k}$ & per unit transportation for forging $k$\\
    $D^i_d$ & discount level $d$ for part $i$, as shown in Fig.~\ref{fig_discounts}\\
    $D^i$ & discount level calculated for part $i$\\
    $|D^i|$ & total number of discount levels\\
    $D^k_d$ & discount level $d$ for forging $k$\\
    $|D^k|$ & total number of discount levels\\
    $\left(Q_{(d-1)}^i, Q_{d}^i\right]$ & quantity interval for discount level $D^i_d$\\
    $\left(Q_{(d-1)}^k, Q_{d}^k\right]$ & quantity interval for discount level $D^k_d$\\
    $\mathbb{M}$ & a very large number\\
    $\mathbb{E}$ & a very small number\\
    \hline
    \multicolumn{2}{l}{\textbf{Variable:}}\\
    $z_{k}$ & 1 if forging $k$ is selected in solution, i.e., consolidated set contains forging $k$, otherwise 0\\
    \hline
    \multicolumn{2}{l}{\textbf{Auxiliary variables:}}\\
    $v_{i}$ & continuous variable to calculate per unit variable cost of part $i$\\
    $x_{ik}$ & indicator variable; 1 if forging $k$ is used to manufacture part $i$ otherwise 0\\
    $u_{dk}$ & indicator variable; 1 if forging $k$ is purchased at discount level $d$ otherwise 0\\
    $y_{ikd}$ &binary variable used for linearisation\\
    $w_{ikd}$ & binary variable used for linearisation\\
    \hline
\end{tabular}
\end{table}

\paragraph{Objective function:}
The procurement cost of parts in the supply chain consists of the sum of forging cost, machining cost and associated inventory holding costs. So, the optimisation problem for procurement cost optimisation through forging consolidation is given as:
\begin{equation}
\label{eq_obj}
\begin{array}{l}
    \min \quad C_M + C_F + C_I,
\end{array}
\end{equation}
where $C_M, C_F$ and $C_I$ are machining costs, forging costs and inventory holding costs, respectively.

The costs of machining $C_M$ to manufacture a given order of parts is the sum of fixed cost and variable cost, which depend on number of units ordered, unit cost of part and unit transportation cost. The machining cost also depends on the forging used to manufacture a part. There can be multiple ways to manufacture a part from different forgings. A forging with minimum machining cost to manufacture a part is selected from consolidated set of forgings. So, the cost of machining can be calculated as given below:
\begin{equation}
    \label{eq_cm}
    \begin{array}{ll}
        C_M &= \sum_i \left[ CMF_i + \left( 1 - D^i \right) \times M_i \times v_i \right]
    \end{array}
\end{equation}
where $CMF_i, D^i, M_i$ and $v_i$ are fixed cost, discount level, ordered quantity and variable cost for part $i$, respectively. $D^i$ can be pre-computed using discount levels $D^i_d$ and quantity intervals $Q_{d}^i$ (as defined in Fig.~\ref{fig_discounts}) because both are available before solving the problem.

Similarly, the cost of forging is sum of fixed cost and variable cost which depend on number of units ordered, unit cost of forging and unit transportation cost of forging. But number of forgings depend on number of parts ordered, and requirement of forging $k$ is calculated using $\sum_i L_{ik} \times M_i \times x_{ik}$, where $L_{ik}$ is number of forgings $k$ required to manufacture one unit of part $i$ and $x_{ik}$ is an indicator variable which indicates if forging $k$ is used to manufacture part $i$. So, assuming $CFF_k$, $CFU_k$ and $CFT_k$ denote fixed cost, unit cost and unit transportation cost, respectively, associated with ordering forging $k$ then forging cost $C_F$ is calculated as below.
\begin{equation}
    \label{eq_cf}
    \begin{array}{l}
        C_F = \sum_{k} z_k \times [ CFF_k + \sum_i L_{ik} \times M_i \times x_{ik} \times \sum_{d} \left( CFU_k + CFT_k \right) \times  \left( 1 - D^k_d\right) \times u_{dk}],
    \end{array}
\end{equation}

where $D^k_d$ and $u_{dk}$ are discount level and corresponding indicator variable.

The inventory cost $C_I$ results from the need to keep an inventory of forgings to manufacture $P_i$ parts. It is sum of cost for purchasing inventory, which is like $C_F$ except fixed cost, and cost of holding inventory. So, assuming $CFH_k$ be unit holding cost for forging $k$, $C_I$ is given below.
\begin{equation}
    \label{eq_ci}
    \begin{array}{l}
        C_I = \sum_{i,k} CFH_k \times L_{ik} \times P_i \times x_{ik} + \sum_{k} z_k \times 
        \left[ \sum_i L_{ik} \times P_i \times x_{ik} \times \sum_{d} \left( CFU_k + CFT_k \right) \times \left( 1 - D^k_d \right) \times u_{dk} \right].
    \end{array}
\end{equation}

\paragraph{Constraints:} 
The problem requires that there should be at least one way to manufacture each part, which can be added as given below.
\begin{equation}
    \label{eq_part_exists}
    \begin{array}{l}
        v_i \ge \mathbb{E}, \quad \forall i,
    \end{array}
\end{equation}
where $v_i$ is per unit variable cost for part $i$ and $\mathbb{E}$ is a very small number (included to avoid strict inequalities), and this inequality ensures that part $i$ is being manufactured. To ensure $v_i$ takes minimum value, i.e., part $i$ is being machined from the cheapest forgings, we further add following constraints.
\begin{equation}
\label{eq_min_forging}
\begin{array}{l}
    v_i  \ge z_k \times \left(CMU_{ik} + CMT_{ik}\right) - \mathbb{M} \times 
    \left( 1- x_{ik} \right), \quad \forall i, k\\
    \sum_k x_{ik} = 1, \quad \forall i,\\
    x_{ik} \le z_k, \quad \forall i, k,
\end{array}
\end{equation}
where $CMU_{ik}$ and $CMT_{ik}$ are per unit machining cost and transportation cost for manufacturing part $i$ from forging $k$, $\mathbb{M}$ is a very large number and $x_{ik}$ is an auxiliary indicator variable which helps to find minimum cost forging for the part. First and second part ensure that machining cost is equal to the minimum of different ways to manufacture the part, and third part of (\ref{eq_min_forging}) ensures that non-zero minimum value is selected.
	
Constraints related to discounts for forgings, i.e., economies-of-scale for forgings are given below, where $d=0,1,2,...,\vert D^k\vert-1$ are discount levels for forging $k$, $u_{dk}$ is an indicator variable that indicates if forging $k$ gets discount level $d$, and $Q_{d}^k$ represents quantity intervals to calculate discounts, as explained in Fig.~\ref{fig_discounts}. The quantity discounts are calculated on forgings required to meet order and inventory requirements, as given below.
\begin{equation}
    \label{eq_one_dk}
    \begin{array}{l}
    \sum_d 	u_{dk} = 1, \quad \forall k.
    \end{array}
\end{equation}

For $d=1,2,...,\vert D^k\vert - 2$ and $\forall k$, 
\begin{equation}
    \label{eq_discount_forgings1a}
    \begin{array}{l}
        u_{dk} \times \left[ \sum_i L_{ik} \times \left( M_i + P_i\right) \times x_{ik} \right] \le Q_{d+1}^k,\\
        u_{dk} \times Q_{d}^k \le \sum_i L_{ik} \times \left( M_i + P_i \right) \times x_{ik} - \mathbb{E}.
    \end{array}
\end{equation}

For extreme cases of $d=0$ and $d=\vert D^k\vert-1$, and $\forall k$,
\begin{equation}
    \label{eq_discount_forgings1b}
    \begin{array}{l}
        u_{0k} \times \left[ \sum_i L_{ik} \times \left( M_i + P_i \right) \times x_{ik} \right] \le Q_{1k},\\
        u_{(\vert D^k\vert-1)k} \times Q_{(\vert D^k\vert - 1)k} \le \sum_i L_{ik} \times \left( M_i + P_i \right) \times x_{ik} - \mathbb{E}.
    \end{array}
\end{equation}

Equation (\ref{eq_one_dk}) ensures that only one discount level is applicable, and inequalities (\ref{eq_discount_forgings1a}) and (\ref{eq_discount_forgings1b}) force that discount level $D^k_d$ is applicable, i.e., $u_{dk}=1$ when required quantity of forging $k$ is in semi-closed interval $\left(Q_{d}^k, Q_{(d+1)}^k\right]$.

The objective function and constraints have two non-linear terms as $z_k \times x_{ik} \times u_{dk}$ and $x_{ik} \times u_{dk}$ which can be simplified into linear terms to make the problem easier to solve. Hence, the following new variables are introduced for linearisation as $y_{ikd} = z_k \times x_{ik} \times u_{dk}$ and $w_{ikd} = x_{ik} \times u_{dk}$, and corresponding constraints are given below.
\begin{equation}
    \label{eq_discount_Yikd}
    \begin{array}{ll}
        y_{ikd} &\le z_k,\\
        y_{ikd} &\le x_{ik},\\
        y_{ikd} &\le u_{dk},\\
        y_{ikd} &\ge z_k + x_{ik} + u_{dk} - 2, \quad \forall i, k, d
    \end{array}
\end{equation}
\begin{equation}
    \label{eq_discount_Wikd}
    \begin{array}{ll}
        w_{ikd} &\le x_{ik},\\
        w_{ikd} &\le u_{dk},\\
        w_{ikd} &\ge x_{ik} + u_{dk} - 1, \quad \forall i, k, d.
    \end{array}
\end{equation}

So, simplifying constraint (\ref{eq_discount_forgings1a}) and (\ref{eq_discount_forgings1b}) using linearisation variables, we get following updated constraints.

For $d=1,2,...,\vert D^k\vert - 2$ and $\forall k$, 
\begin{equation}
\label{eq_discount_forgings2a}
\begin{array}{l}
    \sum_i L_{ik} \times \left( M_i + P_i \right) \times w_{ikd} \le Q_{d+1}^k,\\
    u_{dk} \times Q_{d}^k \le \sum_i L_{ik} \times \left( M_i + P_i \right) \times x_{ik} - \mathbb{E}.
\end{array}
\end{equation}

For extreme cases of $d=0$ and $d=\vert D^k\vert-1$, and $\forall k$,
\begin{equation}
    \label{eq_discount_forgings2b}
    \begin{array}{l}
        \sum_i L_{ik} \times \left( M_i + P_i \right) \times w_{ik0} \le Q_{1k},\\
        u_{(\vert D^k\vert-1)k} \times Q_{(\vert D^k\vert - 1)k} \le \sum_i L_{ik} \times \left( M_i + P_i \right) \times x_{ik} - \mathbb{E}.
    \end{array}
\end{equation}

\paragraph{Optimisation problem:}
The multi-tier material consolidation optimisation problem can be obtained by substituting values of $C_M, C_F$ and $C_I$ into objective function (\ref{eq_obj}), and simplifying the non-linear terms using the linearisation variables, yielding an MILP formulation as given below.
\begin{equation}
    \label{eq_obj2}
    \begin{array}{l}
        \min \quad C_M + C_F + C_I,\\
         =  \sum_i \left[ CMF_i + \left( 1 - D^i \right) \times M_i \times v_i \right]\\ + \sum_{k} z_k \times \left[ CFF_k + \sum_{d} \left( CFU_k + CFT_k\right) \times \left( 1 - D^k_d \right)  \times \sum_i L_{ik} \times M_i \times x_{ik} \times u_{dk} \right]\\
         + \sum_{i,k} CFH_k \times L_{ik} \times P_i \times x_{ik} + 
        \sum_{k} z_k \times \left[ \sum_{d} \left( CFU_k + CFT_k \right) \times \left( 1 - D^k_d \right) \times \sum_i L_{ik} \times P_i \times x_{ik} \times u_{dk} \right]\\
        = \sum_i \left[ CMF_i + \left( 1 - D^i \right) \times M_i \times v_i \right]\\ + \sum_{k} z_k \times \left[ CFF_k + \sum_{d} \left( CFU_k + CFT_k\right) \times \left( 1 - D^k_d\right)  \times \sum_i L_{ik} \times \left(M_i + P_i\right)\times x_{ik} \times u_{dk} \right]\\
         + \sum_{i,k} CFH_k \times L_{ik} \times P_i \times x_{ik}\\
        =  \sum_i \left[ CMF_i + \left( 1 - D^i \right) \times M_i \times v_i \right] + \sum_{i,k} CFH_k \times L_{ik} \times P_i \times x_{ik}\\ + \sum_{k} \left[ z_k \times CFF_k + \sum_{d} \left( CFU_k + CFT_k\right) \times \left( 1 - D^k_d\right) \times \sum_i L_{ik} \times \left(M_i + P_i\right)\times y_{ikd}\right].
    \end{array}
\end{equation}
Subject to constraints (\ref{eq_part_exists}), (\ref{eq_min_forging}), (\ref{eq_one_dk}), (\ref{eq_discount_Yikd}), (\ref{eq_discount_Wikd}), (\ref{eq_discount_forgings2a}) and (\ref{eq_discount_forgings2b}), where variable $z_k$ and auxiliary variables $x_{ik}, u_{dk}, y_{ikd} \text{ and } w_{ikd}$ are binary.

\subsection{Solution approach}
\label{sec_solution}
In this section, we discuss the methodology used to solve the consolidation problem, which is based on efficient problem formulation and the use of exact methods to obtain an optimal solution, as discussed below.

\subsubsection{Model development}
\label{subsec_milp}
For the multi-tier material consolidation problem, as discussed in the previous subsection, we developed an MILP model using the following mathematical techniques for simplification.


\paragraph{Linearisation:} The consolidation problem results in a cubic integer programming problem due to the interaction of two tiers and the exploitation of economies-of-scale. Since non-linear problems are more complex than MILP, we simplified the model using linearization, a process that converts non-linear terms into linear terms by introducing additional auxiliary variables. For example, $x_1$ and $x_2$ are two binary variables and we want to linearise non-linear term $x_1\times x_2$ then we introduce a new binary variable $z=x_1\times x_2$ such that,
	\begin{equation}
		\label{eq_linearise1}
		z \le x_1, \qquad z \le x_2,
	\end{equation}
	\begin{equation}
		\label{eq_linearise2}
		z \ge x_1 + x_2 -1,
	\end{equation}
	where (\ref{eq_linearise1}) ensures that $z$ is 0 when any of $x_1$ or $x_2$ is 0 and (\ref{eq_linearise2}) ensures that $z$ is 1 when both $x_1$ and $x_2$ are 1. Similarly, for $n$ binary variables $x_1,x_2,...,x_n$, non-linear term $\prod_{i=1}^{n} x_i$ can be linearised by introducing $z=\prod_{i=1}^{n} x_i$ as given below.
	\begin{equation}
		\label{eq_linearise3}
		z \le x_i, \quad \forall i,
	\end{equation}
	\begin{equation}
		\label{eq_linearise4}
		z \ge \sum_i x_i - (n-1).
	\end{equation}
	
	Now suppose, $v$ is a continuous variable and $u$ is a binary variable then to linearise $v\times u$, we introduce a continuous variable $s=v\times u$ such that,
	\begin{equation}
		\label{eq_linearise5}
		\begin{array}{c}
			s \le u\times M,\\
			s \le v,\\
			s \ge v - (1-u)\times M,\\
			s \ge 0.
		\end{array}
	\end{equation}
	Here, the first and last inequalities ensure that $s$ is 0 when $u$ is 0. Second inequality ensures that $s$ is upper bounded by $v$ and third inequality ensures that $s$ is lower bounded by $v$ when $u$ is 1, i.e., $s=v$. 
	
\paragraph{Pre-computations:} The concept of pre-computation involves the pre-calculation of values needed in model development, where feasible, to simplify the optimisation process. For instance, in our proposed model, part discounts can either be determined by the model through optimisation or pre-computed because all the necessary information for calculating part discounts is provided. In contrast, discounts for forgings cannot be pre-calculated since the required information is not initially available but rather dependent on the parts.
	
\subsubsection{Exact optimisation}
\label{subsec_exact_method}
The selection of an optimisation method to solve a problem depends on a variety of factors, including complexity of the problem and exact solution requirements. Exact methods generally provide optimal solution to the problem but can take longer to solve in some cases or maybe infeasible (\cite{tavana2022comprehensive,chauhan2022trolley}). In contrast, approximation methods such as heuristics and meta-heuristics can often provide faster solutions but do not guarantee optimality. The consolidation problem can be solved using exact methods as these methods can provide optimal solution for small to large-scale problems.
	
Utilising an exact method involves choosing a modeling language and selecting a solver library. Modelling languages themselves do not solve the problem; instead, they provide an interface where different solvers can be integrated without altering the code. Modelling languages themselves do not solve the problem rather provide interface where different solvers can be plugged without changing the code. Hence, the time to solve an optimisation problem can be divided into two parts: time to generate model using the modelling language and time to solve the problem using the method. In some cases, modelling language can take more than half of the total time to just generate a model (\cite{Lee2020}). Solver libraries offer methods that solve the optimisation problem, such as Gurobi and CPLEX. Please refer to \cite{Anand2017,Lee2020} for a comparative study of optimisation solvers.

	\begin{figure*}[htb!]
		\centering
		\begin{subfigure}[t]{\textwidth}	
			\centering
			\includegraphics[width=0.7\textwidth]{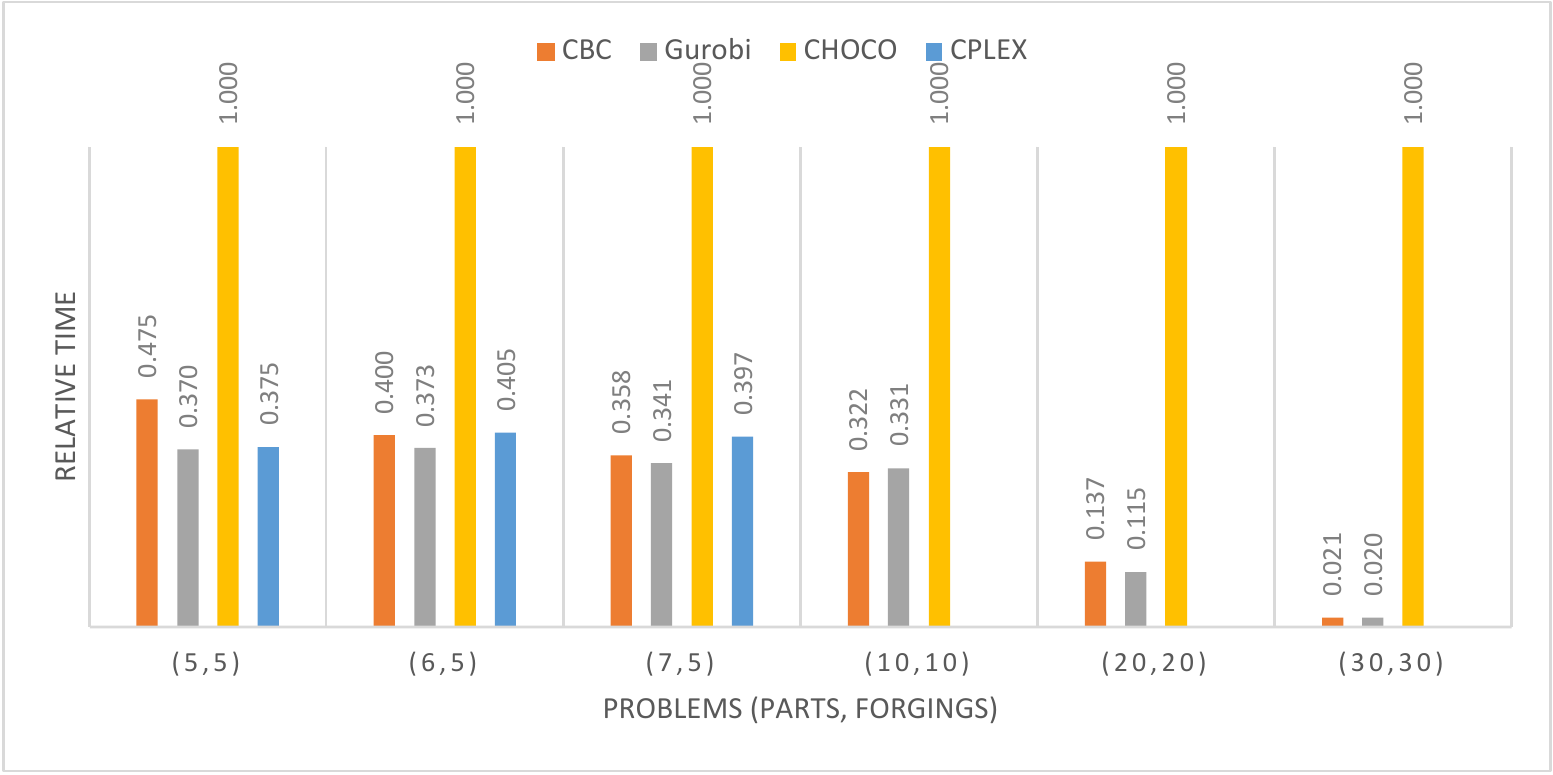}
			\caption{Comparison of different solvers.}
			\label{subfig_Solvers_comparison}
		\end{subfigure}%
		
		\begin{subfigure}[t]{\textwidth}			
			\centering
			\includegraphics[width=0.7\textwidth]{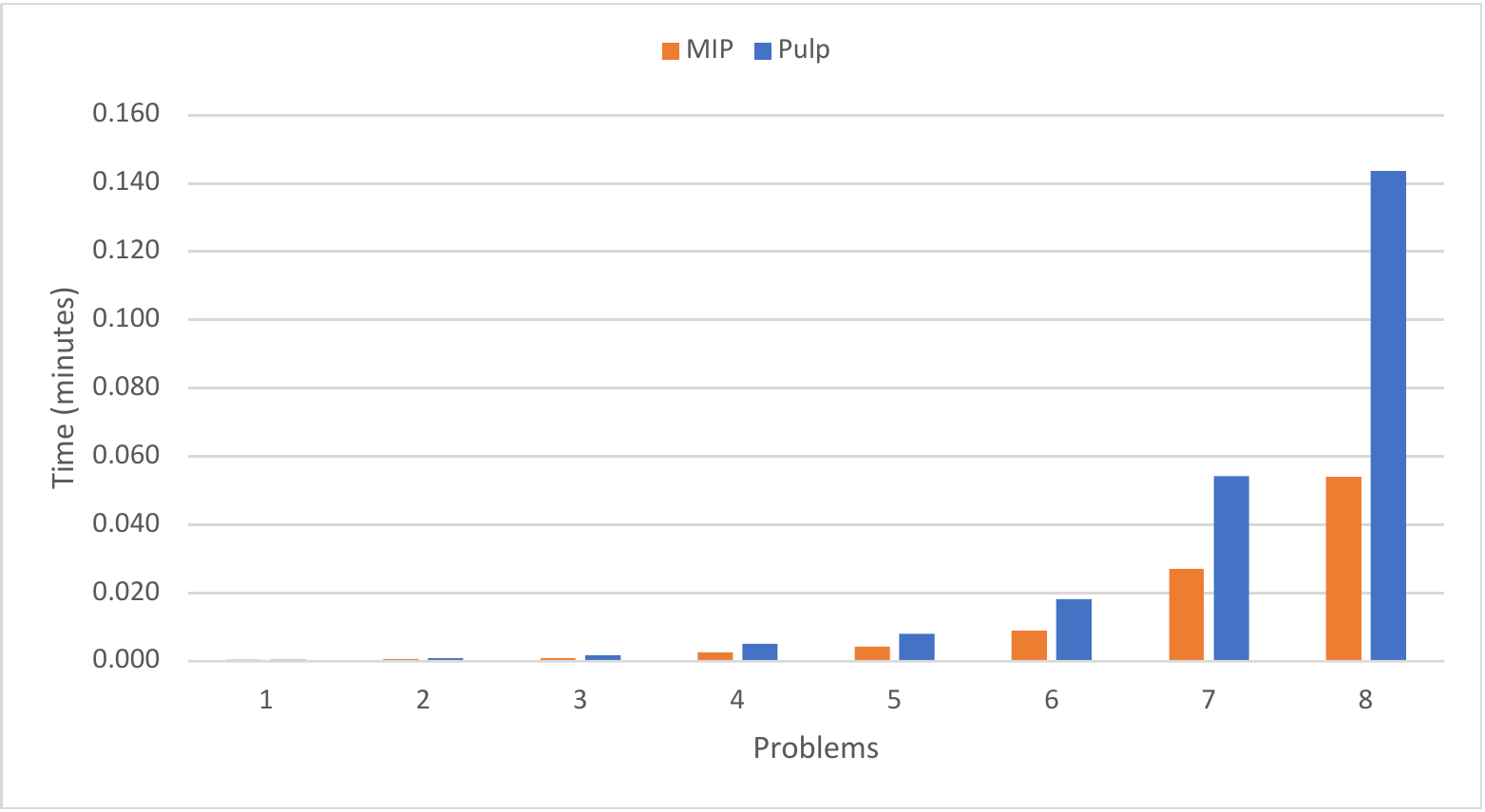}
			\caption{Comparison of Python MIP and Pulp modelling languages.}
			\label{subfig_ModellingL_comparison}
		\end{subfigure}	
		\caption{Comparison of solver libraries and modelling languages.}
		\label{fig_libraries}
	\end{figure*}
	
To solve the multi-tier material consolidation problem, we explored freely available libraries such as CBC, Gurobi (with a free academic license), and CPLEX (free for smaller problems) as solvers. We also utilised Python MIP and Pulp as modelling libraries. We present comparative study of different modelling languages and solvers in Fig.~\ref{fig_libraries}, which presents time (in minutes) to solve different test problems listed in Table~\ref{tab_test_problems}. For CPLEX, results are presented only on first three problems because its free access is limited smaller problems. Sub-figure~\ref{subfig_Solvers_comparison} presents comparative study of different solvers and as it is clear from the figure, Gurobi is the best solver for the consolidation problem and takes significantly lesser time to solve the problems. CBC, the free solver also performs better than Choco for the test problems. Sub-figure~\ref{subfig_ModellingL_comparison} presents comparative study of modelling languages and shows that Python MIP is better than Pulp library to solve the problem under consideration.

\subsection{Experimental results}
\label{subsec_results}
In this subsection, we discuss experimental setup, data and results, as given below.
	
\subsubsection{Data and experimental settings}
\label{subsubsec_settings}
We solve the multi-tier material consolidation problem for an aerospace manufacturing company using representative data.

The order for each part is generated using a random integer drawn from a uniform distribution between 100 and 500, and inventory requirements of each part are generated by the same process to be between 10 and 50. One unit of each forging can manufacture up to three units of parts. Fixed ordering costs for parts and forgings are generated randomly between £1000 and £5000. Machining costs for each part are generated randomly between £10 to £40 and unit transportation costs are generated between £1 to £5. Each part has up to two options to manufacture from forgings. Holding cost per unit of each forging is randomly generated between £15 and £40, per unit forging cost is between £10 and £40, and per unit transportation cost for each forging is between £5 and £25. Although the model can take different discount levels for each part and forging, for simplification of experiments, all parts have same discount levels of 0\%, 5\% and 10\% in the intervals [0, 250], (250, 400] and (400, $\infty$), respectively, and all forgings also have same discount levels of 0\%, 5\% and 10\% in the intervals [0, 250], (250, 400] and (400, $\infty$), respectively. Test problems of small to large sizes are also generated to work with the model as presented in Table~\ref{tab_test_problems}.
	
The experiments are coded in Python programming language, using Python MIP and Pulp as modelling libraries, and CBC, Gurobi, CPLEX and Choco are explored as solvers. Python MIP and Gurobi are used as modelling language and solver libraries, respectively, for reporting the results as they outperform others on the problem, as discussed in \ref{subsec_exact_method}. All the experiments are executed on a MacBook Pro (16GB RAM, 256 SSD, 2.5 GHz Dual-Core Intel Core i7). The code and simulated data are available on GitHub (refer to Section~\ref{sec_code}).
	
\subsubsection{Results}
\label{subsubsec_results}
\begin{figure*}[htb!]
    \centering
    \includegraphics[width=0.6\linewidth]{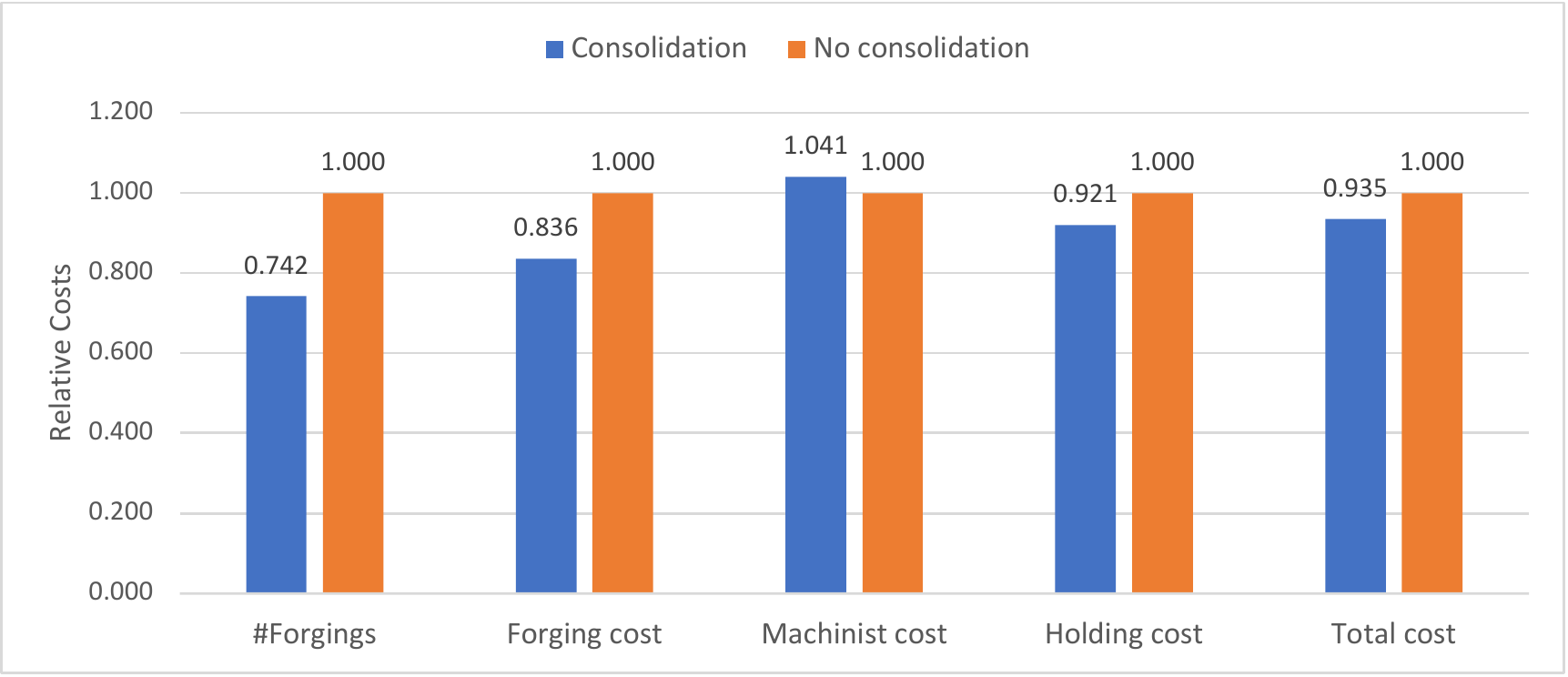}
    \caption{Multi-tier material consolidation case study (500 parts/500 forgings)}
    \label{fig_FC}
\end{figure*}

Fig.~\ref{fig_FC} presents results for the representative case study with 500 parts and 500 forgings. The model takes two minutes to produce the optimal solution for the case study. The multi-tier material consolidation results are benchmarked against results without consolidation and are presented as ratio of values with consolidation to values without consolidation. So, in Fig.~\ref{fig_FC}, values below one show a decrease and values above one show increase in value. For example, a value of 0.74 means that value is reduced by 26\%. As it is clear from the figure, the final consolidated set has around 74.2\% forgings of the given forging set, which shows a 25.8\% consolidation, i.e., the final consolidated set has 25.8\% less forgings than the given set of forgings. The consolidation depends on number of machining options available to manufacture each part from the forgings (please refer to next subsection to study effect of machining options on consolidation). Looking at different costs associated with the problem, it is observed that forging cost is significantly reduced. The decrease in forging cost depends on the fixed costs (ordering costs) associated with each forging (please refer to next subsection to study the effect of ordering cost of forging on consolidation). However, the machining cost, as expected, has slightly gone up because of the trade-off between forging and machining tiers as consolidated forgings require more machining time for manufacturing. Interestingly, holding cost for inventory of forgings has also come down with forging consolidation due to the model which optimises for the overall procurement cost.
	
It is observed that the resulting trade-off between forging cost and machining cost reduces the overall procurement cost in the supply chain. The reduction in procurement cost is a result of coordinating both tiers involved in the trade-off, which will be discussed in the following subsection.

\begin{table}[htb!]
    \centering
		\caption{Study of different test problems}
		\label{tab_test_problems}
		\begin{tabular}{rrrrrrrrr} \hline
			\multicolumn{1}{l}{\textbf{Problem}} & \multicolumn{1}{l}{\textbf{Parts}} & \multicolumn{1}{l}{\textbf{Forgings}} & \multicolumn{1}{l}{\textbf{Consolidated}} & \multicolumn{1}{l}{\textbf{\begin{tabular}[c]{@{}l@{}}Forging\\ Cost\end{tabular}}} & \multicolumn{1}{l}{\textbf{\begin{tabular}[c]{@{}l@{}}Machining\\ Cost\end{tabular}}} & \multicolumn{1}{l}{\textbf{\begin{tabular}[c]{@{}l@{}}Holding\\ Cost\end{tabular}}} & \multicolumn{1}{l}{\textbf{\begin{tabular}[c]{@{}l@{}}Total\\ Cost\end{tabular}}} & \multicolumn{1}{l}{\textbf{\begin{tabular}[c]{@{}l@{}}Time\\(m)\end{tabular}}} \\ \hline
			1 & 5 & 5 & 0.4000 & 0.7751 & 1.0546 & 0.8825 & 0.9029 & 0.001 \\
			2 & 10 & 5 & 0.8000 & 0.9394 & 1.0202 & 0.9792 & 0.9848 & 0.001 \\
			3 & 10 & 10 & 0.7000 & 0.7448 & 1.0805 & 0.9003 & 0.8989 & 0.002 \\
			4 & 20 & 15 & 0.7333 & 0.9037 & 1.0286 & 0.9575 & 0.9652 & 0.004 \\
			5 & 25 & 20 & 0.6000 & 0.8370 & 1.0171 & 0.8992 & 0.9296 & 0.008 \\
			6 & 45 & 25 & 0.8800 & 0.8581 & 1.0572 & 0.8406 & 0.9569 & 0.009 \\
			7 & 70 & 50 & 0.8400 & 0.9090 & 1.0467 & 0.9564 & 0.9780 & 0.028 \\
			8 & 100 & 70 & 0.8571 & 0.8832 & 1.0411 & 0.9480 & 0.9643 & 0.055 \\
			9 & 100 & 100 & 0.7400 & 0.8501 & 1.0353 & 0.9328 & 0.9397 & 0.079 \\
			10 & 200 & 150 & 0.8000 & 0.8572 & 1.0569 & 0.9273 & 0.9554 & 0.278 \\
			11 & 200 & 200 & 0.7200 & 0.8489 & 1.0376 & 0.9404 & 0.9407 & 0.322 \\
			12 & 500 & 250 & 0.8880 & 0.8831 & 1.0417 & 0.9009 & 0.9637 & 1.011 \\
			13 & 500 & 500 & 0.7420 & 0.8360 & 1.0410 & 0.9206 & 0.9348 & 1.978 \\
			14 & 1000 & 700 & 0.8114 & 0.8726 & 1.0434 & 0.9322 & 0.9589 & 5.966 \\
			15 & 1000 & 1000 & 0.7460 & 0.8514 & 1.0368 & 0.9303 & 0.9421 & 7.996 \\
			16 & 1500 & 1000 & 0.8290 & 0.8685 & 1.0454 & 0.9120 & 0.9579 & 12.386 \\
			17 & 1500 & 1500 & 0.7493 & 0.8420 & 1.0454 & 0.9188 & 0.9401 & 18.771 \\
			18 & 2000 & 1800 & 0.7567 & 0.8492 & 1.0485 & 0.9130 & 0.9460 & 32.595 \\
			19 & 2000 & 2000 & 0.7385 & 0.8467 & 1.0412 & 0.9294 & 0.9408 & 40.786 \\
			20 & 3000 & 2500 & 0.0426 & 0.8559 & 1.0404 & 0.9209 & 0.9480 & 106.602 \\ \hline
		\end{tabular}
	\end{table}
	
	Table~\ref{tab_test_problems} presents the scale of forging consolidation test problems that we have attempted. Like Fig.~\ref{fig_FC}, values in the table are relative to values without consolidation and presented as a ratio of values with consolidation to without consolidation. The proposed model can solve all problem instances to optimality, and we observe results similar to Fig.~\ref{fig_FC}. All problems show corresponding decrease in forging costs at the expense of a slight increase in the correspondence machining cost. The optimum balance in the trade-off between forging and machining cost, also helps to reduce the overall procurement cost significantly for all the test problems. Interestingly, the consolidation approach is also able to bring down the holding cost. The last column of the table displays computational time in minutes and provides sensitivity analysis against problem size. As it can be observed from the table, the model is able to solve small to large-scale problems in reasonable time.
	
\subsubsection{Sensitivity analysis}
\label{subsubsec_sensitivity}
Here, we study effect of fixed forging cost, machining options, forging discounts, and holding cost on forging consolidation. For simplicity of experiments, sensitivity analysis is performed on a test problem of 100 parts and 100 forgings, and the problem is solved using Python MIP as modelling library and Gurobi as solver library.
	
\begin{figure*}[htb!]
    \centering
    \includegraphics[width=0.8\linewidth]{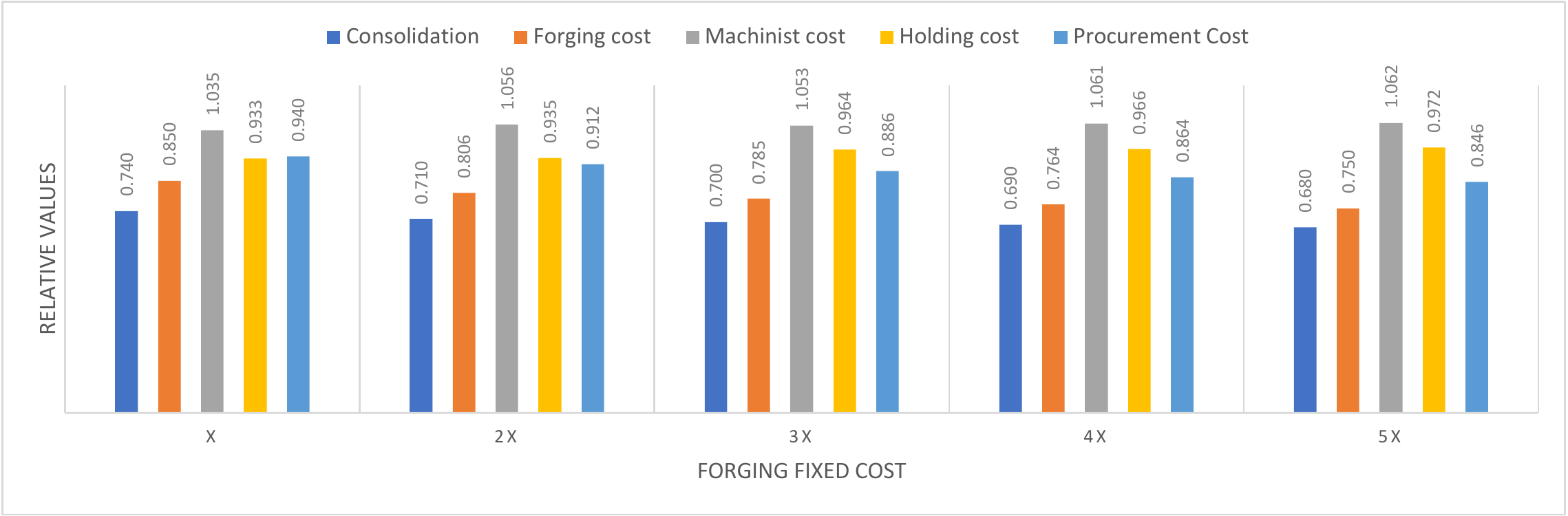}
    \caption{Study of effect of forging ordering/fixed cost on the consolidation (100 parts/100 forgings)}
    \label{fig_Forging_Ordering_Cost}
\end{figure*}
Fig.~\ref{fig_Forging_Ordering_Cost} depicts impact of fixed/ordering cost on the forging consolidation problem. In this study, we increase fixed cost by a factor of one to five times and study impact of this increase on the problem. It is clear that an increase in fixed/ordering forging cost monotonically increases the consolidation (i.e., the consolidated set has a smaller number of forgings), monotonically decreases the consolidation cost, monotonically increases holding cost, and monotonically decreases overall procurement cost in the supply chain. Although, an increase in fixed cost of forgings increases holding cost, holding cost is still lesser than non-consolidated case. Additionally, an increase in fixed cost of forgings increases machining cost but there is no steady increase in machining cost.
	
\begin{figure*}[htb!]
    \centering
    \includegraphics[width=0.8\linewidth]{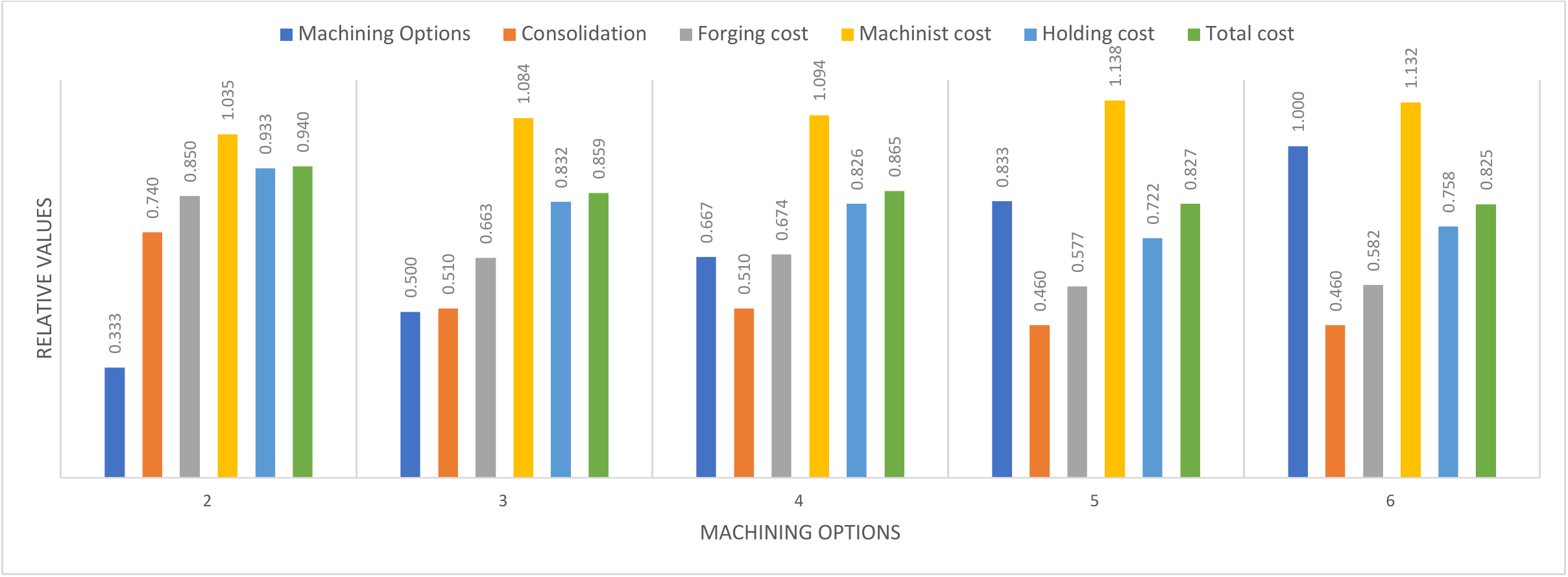}
    \caption{Study of machining options (100 parts/100 forgings)}
    \label{fig_Machining_Options}
\end{figure*}
Fig.~\ref{fig_Machining_Options} presents impact of number of machining options (i.e., number of ways to manufacture a part from different forgings) available for each part on the consolidation. We consider up to two, three, four, five and six options for each part. For ease of interpretation, machining options are represented as the ratio of machining options divided by the maximum options considered, i.e, six. From the figure, it is clear that with an increase in machining options, in general, there is an associated increase in consolidation, decrease in forging cost, decrease in holding cost and decrease in the procurement cost. This is because increase in machining options means that there are more ways to manufacture the same part and thus more opportunities for consolidation. There is also slight increase in machining cost. Further analysis of actual values, shows that with an increase in machining options there is a monotonic increase in consolidation, monotonic decrease in forging cost and procurement cost. This difference in actual costs and relative costs (from the figure) is due to the fact that an increase in machining options affects both solutions, i.e., solutions with and without consolidation.

\begin{figure*}[htb!]
    \centering
    \includegraphics[width=0.8\linewidth]{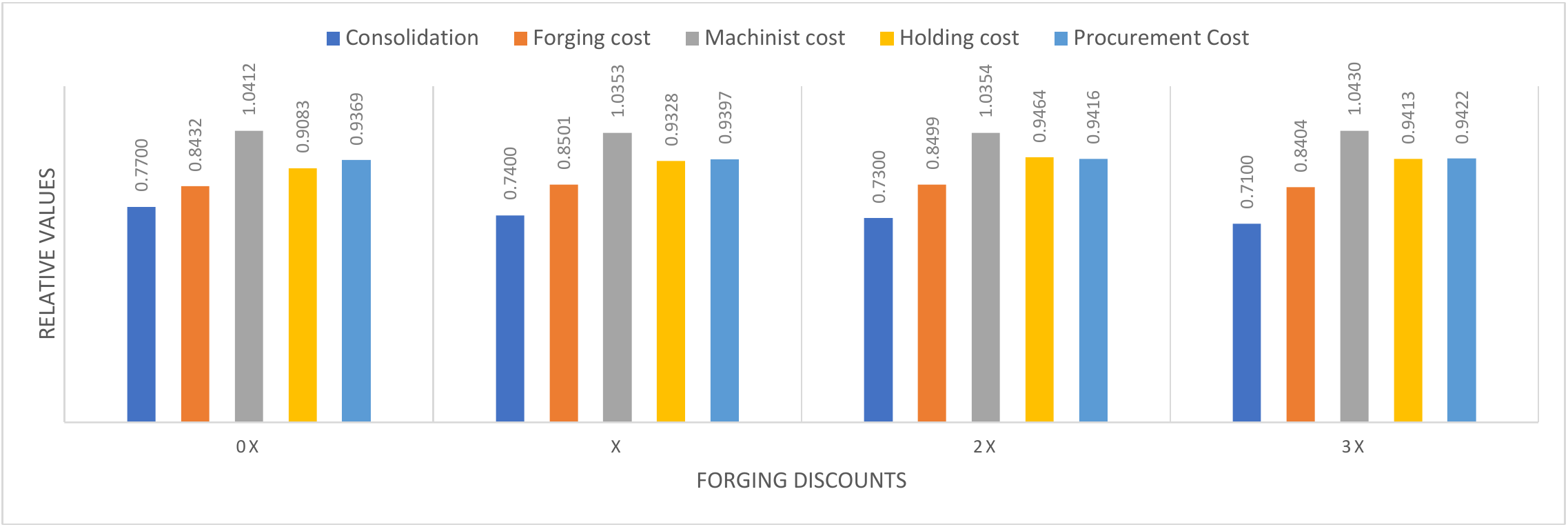}
    \caption{Sensitivity analysis using forging discounts (100 parts/100 forgings).}
    \label{fig_Forging_Discounts}
\end{figure*}

The effect of forging cost discounts on consolidation is presented using Fig.~\ref{fig_Forging_Discounts}. As it is clear from the figure, with an increase in discounting there is monotonic increase in consolidation. This is because discounting helps to exploit economies-of-scale as consolidated set of forgings has fewer forgings with more demands than non-consolidated set which has more forgings with lesser demands. Moreover, an increase in forging discount shows a slight increase in total cost. At first, this looks strange but can be explained from actual cost values (which shows decrease in cost) and the fact that non-consolidation solution also benefits from increase in discounts. However, there is no clear observable patterns in forging cost, machining cost and holding cost. This is because an increase in discounts also impacts the non-consolidation based solution. In addition, forging consolidation is effective in both scenarios, i.e., with and without discounts on forgings, as shown by case `0X' in the figure.

\begin{figure*}[htb!]
    \centering
    \includegraphics[width=0.8\linewidth]{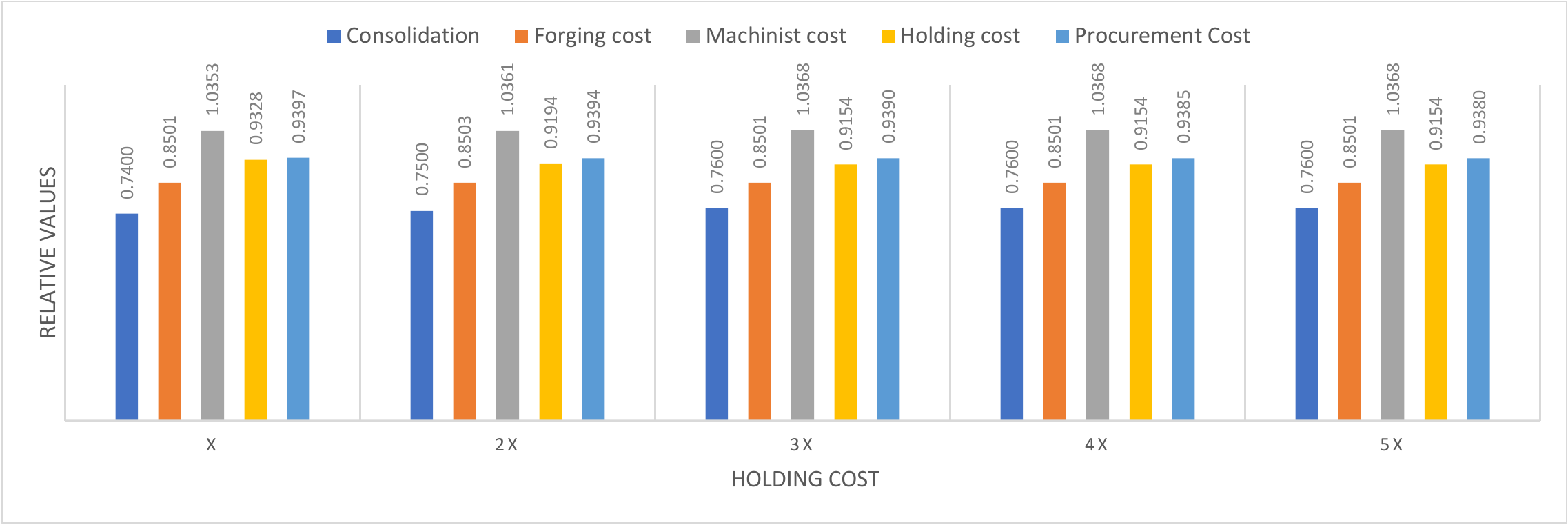}
    \caption{Sensitivity analysis with Holding Cost (100 parts/100 forgings).}
    \label{fig_Holding_Cost}
\end{figure*}

Fig.~\ref{fig_Holding_Cost} presents effect of per unit holding cost on the consolidation. There are slight changes in costs and there is no clearly observable pattern because an increase in per unit holding cost affects both consolidation based solution and non-consolidation based solution. Consolidation looks to decrease with an increase in holding cost but no corresponding increase in forging cost is observed. However, relative holding cost is always less than one, i.e., consolidation always leads to decrease in holding cost.

\section{Conclusion and Discussion}
\label{sec_conclusion}
Whilst consolidation has been studied widely as a sub-field of cooperation and interaction in supply chain management, the exploitation of multi-to-multi relationship between tiers has not been paid much attention, despite being a prominent feature of complex engineered products in sectors such as medical devices and aerospace precision-machined products.
	
In this paper, we developed a formulation of the multi-tier material consolidation problem, and developed a mixed integer linear programming approach validated through an aerospace case study. Using an efficient problem formulation approach, the model is solved using exact optimisation methods to get optimal solution for a range of small to large-scale problems. The resulting formulation is flexible such that quantity discounts, inventory holding and transport costs can be included. Our comparison to benchmark results where no consolidation activity occurs, show that there is indeed a cost trade-off between two tiers, and that its reduction can be achieved using a holistic approach to reconfiguration.
	
From our sensitivity analysis, we observe that several interesting patterns emerge. Consolidation increases, i.e., relative cost decreases with an increase in fixed ordering cost of forgings and with an increased number of machining options for the parts. It is also observed that discounting helps increase consolidation although there is no clear observable pattern in relative costs. Similarly, consolidation reduces relative holding costs for all test problems but there is no clear observable pattern amongst the different types of costs.
	
Moreover, since consolidation leads to a smaller set of forgings both the inventory management process and subsequent supplier selection process is potentially simplified. For inventory management, consolidation means that the decision space for determining appropriate policies for transport and storage is reduced. Similarly, a reduced number of forgings may result in the reduction of forgers required, hence allowing supply chain simplification. Thus, multi-tier material consolidation not only might lead to a reduction in the procurement cost but may encourage more cooperation between tiers and simplification of procurement and inventory management tasks.

\subsection{Limitations and Future Scope}
\label{subsec_conclusion}
This study has several limitations and assumptions which may lead to potential avenues for future research, as discussed below.
\begin{enumerate}
    \item First of these is the assumption on constant and a priori known demand, for which uncertainty handling methods may be considered. This would be especially prudent in cases where machined parts may differ in their demand profiles, perhaps due to spare parts requirements. 
    \item Another assumption we undertook that consolidation does not lead to any negative effects on supply chain relationships, whereas these may need to be considered within a wider frame of reference, in which different assemblies are procured from suppliers whose products are being consolidated.
    \item We also did not consider the risks of consolidation in the supply chain which could be managed by the manufacturer, while allocating orders to suppliers, by approaches, like, multi-sourcing. 
    \item Finally, we have assumed that reconfiguration requirements to produce more quantities of some of the forgings are applied at no additional cost, which may not be the case.
\end{enumerate}
Our future work will study these effects of multi-tier consolidation in the supply chain and further use cases to shed more light on the benefits and drawbacks of our approach.

\section*{Funding}
This research was funded by Aerospace Technology Institute and Innovate UK, the UK’s innovation funding agency, through the ``Digitally Optimised Through-Life Engineering Services" project (113174).

\section*{Acknowledgements}
The authors gratefully acknowledge the editor-in-chief, the associate editor and the anonymous reviewers for their constructive comments for improving the quality of the paper.

\section*{Conflicts of Interest}

The authors declare that they have no conflict of interest.

\section*{Code Availability}
\label{sec_code}
The code and simulated data used in the paper will be released on acceptance at following link:\\ \url{https://github.com/ifm-mag/DOTES-Forging-Consolidation-2023}.

	


\end{document}